\documentclass[9pt,oneside]{amsart}

\usepackage{tikz}
\usetikzlibrary{automata, positioning, arrows}

      \usepackage{amssymb}
\usepackage{amsfonts,amstext}
\usepackage{graphicx}
\usepackage{amsmath,amstext,amssymb,amscd}
\usepackage{verbatim} 
\usepackage{mathtools}% This package allows the use of \begin{comment}
\usepackage{tabularx}
\usepackage{url}
\usepackage{graphicx}
\usepackage{adjustbox}
\usepackage{amssymb,latexsym}
\usepackage{amsmath}
\usepackage{amsthm}
\usepackage{amssymb}

\theoremstyle{remark}

\theoremstyle{definition}

\usepackage{amsthm}
\usepackage{multicol}

      \makeatletter
      \def\@setcopyright{}
      \def\serieslogo@{}
     
      \makeatother

\begin{document}

\author{David McCune}
\address{David McCune, Department of Mathematics and Data Science, William Jewell College, 500 College Hill, Liberty, MO, 64068-1896}
\email{mccuned@william.jewell.edu} 

\author{Lori McCune}
\address{Lori McCune, Department of Computer Science, Mathematics and Physics, Missouri Western State University, 4525 Downs Dr, St Joseph, MO, 64507}
\email{lmccune@missouriwestern.edu}

\title[ The Curious Case of the 2021 Minneapolis Ward 2 City Council Election]{The Curious Case of the 2021 Minneapolis Ward 2 City Council Election}

\begin{abstract} In this article we explain why the November 2021 election for the Ward 2 city council seat in Minneapolis, MN, may be the mathematically most interesting ranked choice election in US history.\end{abstract}



\maketitle

The 2009 ranked choice mayoral election in Burlington, VT, was a controversial election; so controversial, in fact, that in its aftermath the voters of the city decided to repeal the use of ranked choice voting.  From a mathematical perspective, it is probably the most interesting ranked choice election in US history; that is, it \emph{was} the most interesting until  the 2021 Minneapolis, MN, election for the Ward 2 city council seat.  In this article we explain what made the Burlington election so interesting and controversial (our information about this election comes from \cite{FT} and \cite{ON}), and explain why the Minneapolis election might be the new champion of ranked choice mathematical wackiness in the US.

How does ranked choice voting work?  When voters cast a ballot, they do not vote for a single candidate.  Instead, each voter submits a preference ranking of the candidates where one candidate is listed as the voter's first choice, another candidate is listed as the voter's second choice, and so on.  In the Burlington mayoral election there were five candidates: Bob Kiss, Kurt Wright, Andy Montroll, Dan Smith, and James Simpson.  Each voter provided a ranking of these candidates where the ranking was not required to be complete.  For example, a voter could rank Wright first, Montroll second, and leave all other rankings blank.  The winner was determined using a voting method known as \emph{instant runoff voting} (IRV), which works as follows: first, eliminate the candidate with the fewest first place votes.  Each voter that chose this eliminated candidate as their first choice then has their votes transferred to the candidate that is ranked second on their ballots. Repeat this process of elimination and vote-transfer until a candidate has earned a majority of the remaining votes, and that candidate is crowned the winner.

Table \ref{burlington} shows the election data from Burlington after the non-contending candidates Simpson and Smith were eliminated and their votes transferred.  At this point in the instant runoff process, 2043 voters rank Kiss first, Montroll second, and Wright third; 370 rank Kiss first, Wright second, and Montroll third, etc.  The $-$ indicates that no candidate was ranked in this slot.  For simplicity, we combine ballots of the form $A$, $B$, $C$ with ballots of the form $A$, $B$, $-$.  The first-place vote totals are 2982, 2554, and 3297 for Kiss, Montroll, and Wright, respectively.  Thus Montroll is eliminated and 1332 votes are transferred to Kiss, 767 votes are transferred to Wright, and 455 votes are dropped from the election.  After the dust settles, Kiss wins with 4314 votes to Wright's 4064.

What makes this election interesting?  First, each of the three candidates has a legitimate claim to be the election winner. Kiss wins using IRV and thus it makes sense for him to win.  Wright has the most first place votes in Table \ref{burlington} and thus he would win if we used the method of \emph{plurality}, the usual voting method in the US which declares as winner the candidate with the most first-place votes (we note that Wright received the most first place votes when using the full dataset with five candidates, so he truly is the plurality winner of the election).  Montroll wins the election under two other standard voting methods. Note that 4067 voters prefer Montroll to Kiss while only 3477 prefer Kiss to Montroll; similarly 4597 voters prefer Montroll to Wright while only 3668 voters prefer Wright to Montroll.  Since Montroll wins both of his head-to-head matchups, he is the \emph{Condorcet winner} and any Condorcet method would declare him the winner.  Montroll is also the winner under the classical method \emph{Borda count}.  In this points-based scoring method, a first-place vote in Table \ref{burlington} is worth three points (if there are $n$ candidates in an election, a first-place vote is worth $n$ points), a second-place vote is worth two points, and a last-place vote is worth one point.  The reader can check that Montroll is the Borda winner using the data from Table \ref{burlington} (we note that Montroll is still the Borda winner if we use the complete data before eliminating any candidates). Thus, any of the three candidates has a strong and reasonable claim to be the unique winner of the election.
\begin{table}[]
  \centering

  \begin{adjustbox}{width=\textwidth}

\begin{tabular}{l|ccccccccc}
Number of Voters & 2043 & 371 &568 & 1332 & 767 &455&495&1513&1289\\
\hline
1st choice & Kiss & Kiss & Kiss & Montroll & Montroll &Montroll&Wright &Wright &Wright\\
2nd choice & Montroll & Wright & $-$ & Kiss &Wright &$-$&Kiss&Montroll&$-$\\
3rd choice & Wright & Montroll & $-$ & Wright & Kiss &$-$&Montroll&Kiss&$-$\\
\end{tabular}

 \end{adjustbox}
  \caption{The 2009 mayoral election in Burlington, VT, after eliminating the bottom two candidates.  The raw vote data is available from www.preflib.org; this table appears in \cite{ON}.}
  \label{burlington}
\end{table}

This election was interesting for another reason: who would have won the election if Kiss had persuaded more voters to support him?  In particular, what if 300 voters who had voted Wright, Kiss, Montroll and 450 who had voted Wright, $-$, $-$ had instead voted Kiss, $-$, $-$?  Since Kiss won the original election, it seems that gaining more support should only help him. Shockingly enough, gaining this extra support would actually have cost him the election since now Wright is eliminated first and, after the votes are transferred, Kiss would lose to Montroll.  This is an example of a \emph{monotonicity paradox}, where gaining support would cause a winner to lose an election  (an \emph{upward monotonicity paradox}), which occurs in Burlington, or losing support would cause a loser to win an election (a \emph{downward monotonicity paradox}).  Either type of paradox is very rare to observe in practice: Graham-Squire and Zayatz \cite{GZ} studied a large set of American ranked choice elections in search of monotonicity paradoxes, and the Burlington election was the only example found.

%Borda scores:
%Kiss: 14874
%Montroll: 15636
%Wright: 15531

With the Burlington example provided for context, we now describe the 2021 Minneapolis Ward 2 election, which also used IRV.  This election contained 5 candidates: Cam Gordon, Guy Gaskin, Robin Worlobah, Tom Anderson, and Yusra Arab.  Gaskin and Anderson were relatively weak and eliminated first, resulting in the three-candidate data in Table \ref{minneapolis}. In this case the table is a bit misleading: it shows that Arab is the plurality winner because she has the most first-place votes, but Worlobah was the actual plurality winner because she had the most first-place votes prior to the elimination of  Gaskin and Anderson.  Otherwise using Table \ref{minneapolis} yields the same election winner under various voting methods as the actual data with five candidates.

This election demonstrates the same kind of conflict about the winner as the Burlington election.  Worlobah wins under plurality and IRV (Gordon is eliminated first, then Worlobah beats Arab head-to-head), while Arab is the Borda winner.  What about the Condorcet winner? Arab wins a head-to-head matchup against Gordon 4324 to 4098, Gordon wins a matchup against Worlobah 3708 to 3635, and Worlobah wins a matchup against Arab 4056 to 4037. Every candidate loses one matchup and therefore this election has no Condorcet winner, producing a \emph{Condorcet paradox} in which the voters' preferences are non-transitive (the electorate prefers Arab to Gordon, Gordon to Worlobah, and Worlobah to Arab). This is an extremely rare outcome to observe in actual elections: for another project \cite{MM}, we analyzed every American ranked choice election for which we could obtain and process the raw vote data (and in which there were more than two candidates and no candidate received a strong majority of first place votes), resulting in a dataset of more than 200 elections. This Minneapolis election is the only example without a Condorcet winner and thus is the first documented example of this kind of Condorcet paradox in American political elections, as far as we are aware. Since Worlobah is the IRV winner and Gordon beats Worlobah in a head-to-head matchup, Gordon also has a strong claim to be the election winner and, like Burlington, it is difficult to decide which candidate ``should'' be the winner.

\begin{table}[]
  \centering

  \begin{adjustbox}{width=\textwidth}

\begin{tabular}{l|ccccccccc}
Number of Voters & 801 & 1177 &822 & 908& 756 &1572&1299&1088&492\\
\hline
1st choice & Gordon & Gordon & Gordon & Arab & Arab &Arab&Worlobah&Worlobah &Worlobah\\
2nd choice & Arab & Worlobah & $-$ & Gordon &Worlobah &$-$&Gordon&Arab&$-$\\
3rd choice & Worlobah & Arab & $-$ & Worlobah & Gordon &$-$&Arab&Gordon&$-$\\
\hline
\hline

Precincts 1 and 5 & 182 & 279 & 187 & 165 & 140 & 397 & 291 & 242 & 103\\
All Other Precincts &619 & 898 & 635 & 743 & 616 & 1175& 1008 & 846 & 389\\
\end{tabular}

 \end{adjustbox}

  \caption{The 2021 Ward 2 election for Minneapolis city council after eliminating the bottom two candidates. The bottom of the table gives a division of the voters into two sub-electorates. The vote data is available at https://vote.minneapolismn.gov/results-data/election-results/.}
  \label{minneapolis}
\end{table}

The Minneapolis election also demonstrates a downward monotonicity paradox.  If 80 of the voters who cast the ballot Arab, Gordon, Worlobah had instead moved Arab down on the ballot and voted Gordon, Arab, Worlobah, then Worlobah would be the first candidate eliminated and Arab would be the IRV winner.  The 2021 Minneapolis Ward 2 election provides the second documented instance of a monotonicity paradox in the history of American ranked choice elections, and the first documented example of a downward paradox. 

Finally, this election has another interesting feature. Ward 2 has 11 precincts; suppose we divide the electorate into two groups, one of which contains the voters from precincts 1 and 5 (Group 1) and the other which contains all other voters (Group 2).  The vote data for each separate group is displayed at the bottom of Table \ref{minneapolis}.  If we run the IRV algorithm on Group 1 or Group 2 separately, Arab wins each group. However, when we combine the two groups together to recover the original election data, Arab loses.  This is an example of a \emph{consistency paradox}, a voting theory example of Simpson's paradox. If Arab wins the election for Group 1, and then we combine the ballots of Group 1 with another set of ballots for which Arab wins (Group 2), how can she lose the election for the overall set of all ballots? If we are willing to break apart precincts, we can find hundreds of partitions of the electorate into two groups where Arab wins both groups or Gordon wins both groups, yet Worlobah wins the overall election. For both Arab and Gordon,  the whole is less than the sum of the parts in this election.  We note that this same paradox can be demonstrated for the Burlington election, and we are unaware if this has been observed before. We do not know if there exist empirical studies which investigate the frequency of this paradox in American elections, so we are unsure of its frequency in practice.

%Finally, this election has another interesting feature.  Suppose that Ward 2 were divided into two geographic areas, partitioning the voters into the two electorates displayed in Table \ref{minneapolis}.  Under IRV, if we use the ballots from just Electorate 1 then Arab wins; similarly for Electorate 2.  Yet, when the two electorates are combined to obtain the actual vote data, Arab loses.  This is an example of a \emph{consistency paradox}, where the whole is less than the sum of the parts.  If Arab wins in each individual geographic region, why shouldn't she win the election across the entire ward?  We challenge the reader to find a similar division of the voters into two subgroups but where Gordon wins each individual election; for Gordon, the whole is also less than the sum of the parts. We note that this same paradox can be demonstrated for the Burlington election, and we are unaware if this has been observed before. We are unaware of empirical studies that investigate the frequency of this paradox in American elections, so we are unsure of its frequency in practice.

The 2021 Minneapolis election for city council seat in Ward 2 contained three candidates each of whom has a legitimate claim to be the winner, the first known example of an American political election without a Condorcet winner, the first known example of an American political election containing a downward monotonicity paradox (and the second known example of any type of monotonicity paradox), and multiple consistency paradoxes. Surely this election deserves consideration as the most interesting ranked choice election in US history.

\end{document}